\documentclass[twocolumn,prl,aps]{revtex4}

\usepackage{graphicx}
\usepackage{psfrag}
\usepackage{euscript}
\usepackage{verbatim}
\def\fig_width{3. in} % width of single column figure in PR

\def\threej(#1,#2)(#3,#4)(#5,#6){\begin{pmatrix}#1&#3&#5\\#2&#4&#6\end{pmatrix}}
\def\sixj(#1,#2,#3)(#4,#5,#6){\begin{Bmatrix}#1&#2&#3\\#4&#5&#6\end{Bmatrix}}
\def\ninej(#1,#2,#3)(#4,#5,#6)(#7,#8,#9){\begin{Bmatrix}#1&#2&#3\\#4&#5&#6\\#7&#8&#9\end{Bmatrix}}

\def\twobytwo(#1,#2)(#3,#4){\begin{pmatrix}#1&#2\\#3&#4\end{pmatrix}}

\newlength{\defbaselineskip}
\newcommand{\setlinespacing}[1]%
           {\setlength{\baselineskip}{#1 \defbaselineskip}}

\usepackage{amsmath}
\begin{document}
\title{Nonlinear magneto-optical rotation with frequency-modulated light}
%==================================
\author{D. Budker$^{a,b}$, D. F. Kimball$^{a}$, V. V. Yashchuk$^{a}$, and M. Zolotorev$^{c}$} \affiliation {
\\$^a$ Department of Physics,
University of California, Berkeley, CA 94720-7300
\\e-mail: budker@socrates.berkeley.edu
\\$^b$ Nuclear Science Division, Lawrence Berkeley National Laboratory, Berkeley CA
94720
\\$^c$ Center for Beam Physics, Lawrence Berkeley National
Laboratory, Berkeley CA 94720}

\date{\today}

\begin{abstract}
A magnetometric technique is demonstrated that may be suitable for
precision measurements of fields ranging from the sub-microgauss
level to above the Earth field. It is based on resonant nonlinear
magneto-optical rotation caused by atoms contained in a vapor cell
with anti-relaxation wall coating. Linearly polarized,
frequency-modulated laser light is used for optical pumping and
probing. If the time-dependent optical rotation is measured at the
first harmonic of the modulation frequency, ultra-narrow ($\sim$ a
few Hz) resonances are observed at near-zero magnetic fields, and
at fields where the Larmor frequency coincides with half the light
modulation frequency. Upon optimization, the sensitivity of the
technique is expected to exceed $10^{-11} G/\sqrt{Hz}$.
\end{abstract}

%32.80.Bx Level crossing and optical pumping
%07.55.Ge Magnetometers for magnetic field measurements
%95.75.Hi Polarimetry
%42.65.-k Nonlinear optics
%%% ----------------------------------------------------------------------
\maketitle
%%% ----------------------------------------------------------------------

When light near-resonant to an atomic transition propagates
through an atomic medium immersed in a magnetic field, the light
polarization can be affected. For example, when a magnetic field
is applied along the direction of light propagation, there is
light-power-dependent rotation of the polarization plane known as
nonlinear magneto-optical (Faraday) rotation (NMOR). Recently
\cite{ourULTRA}, we observed ultra-narrow ($\sim 1\ {\rm Hz}$)
zero-field resonances in NMOR with rubidium atoms contained in
vapor cells with high-quality anti-relaxation coating
\cite{Ale96}. These resonances arise due to preservation of atomic
polarization over thousands of collisions with the walls of the
cell. The sensitivity of an NMOR-based magnetometer for
sub-microgauss fields could, in principle, exceed $10^{-11}\
G/\sqrt{Hz}$ \cite{ourSens}, approaching the fundamental
shot-noise limit (given the number of atoms in the cell and the
polarization relaxation rate). In the present work, we describe a
new technique that involves both inducing and measuring NMOR with
a single frequency-modulated light beam. Additional ultra-narrow
resonances are observed when the light modulation frequency
coincides with twice the Larmor frequency, 2$\Omega_L$ (and also
with $\Omega_L$, etc.), the use of which can extend the dynamic
range of an NMOR-based magnetometer to above the Earth field range
($\sim 1\ $G) of interest in many applications.

The high sensitivity of NMOR magnetometry is based in part on the
ability of the polarimeter to measure small optical rotation
angles. The sensitivity of polarimeters without modulation is
limited by low-frequency noise. Therefore, the most successful
polarimetric techniques always involve some type of fast
modulation. In our previous work \cite{ourULTRA,ourSens}, we used
polarization-modulation polarimetry, where in addition to the
vapor cell under investigation, a Faraday rotator is inserted
between the crossed polarizer and analyzer. The Faraday rotator
modulates the polarization direction at a frequency of $\sim 1\
$kHz, and the rotation angle is determined by measuring the first
harmonic of the signal from a photodetector in the dark output of
the analyzer. Such a method allows detection of signals at high
frequencies, eliminating sensitivity to excess low-frequency
noise. However, the polarization modulation technique has a number
of shortcomings. Importantly, it is not immune to the problem of
``drifting zero" -- any relative rotation of the polarizer and
analyzer, or a change in the birefringence of optical elements,
etc. is detected as a rotation signal indistinguishable from
rotation produced by the atoms. In order to avoid such problems, a
light frequency modulation technique was introduced in Refs.
\cite{Bar78,Bar88} and was applied to measurements of
\emph{linear} magneto-optical rotation and parity-violating
optical rotation. This technique allowed detection of optical
rotation at the light modulation frequency without introducing
additional optical elements (such as a Faraday modulator) between
the polarizer and analyzer, considerably reducing spurious
rotations. Sensitivity to most spurious rotations (including the
``drift of zero") is further reduced because such effects
generally do not lead to spectral features as sharp as an atomic
resonance. Optical pumping with frequency-modulated light was
studied and applied to $^4{\rm He}$ resonance magnetometers
\cite{Che95,Che96,Gil2001}, but in that work transmission was
monitored, in contrast to the measurements of optical rotation
performed in the present work. The performance of the $^4{\rm He}$
magnetometer is limited by laser intensity noise \cite{Gil2001},
which is not expected to limit the performance of NMOR-based
magnetometers.

\begin{figure}
\includegraphics[scale=.5]{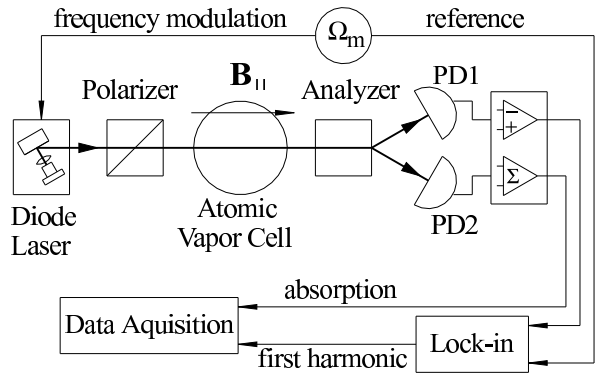} \caption{Simplified
schematic of the apparatus.} \label{FM_Appar}
\end{figure}

A simplified diagram of the experimental apparatus is shown in
Fig. \ref{FM_Appar}. A paraffin-coated vapor cell (diameter 10 cm)
containing an isotopically enriched sample of Rb atoms
($\approx94\%$ of $^{87}{\rm Rb}$; atomic density $\approx
7\times10^9~{\rm cm^{-3}}$ at 20$^\circ$C) is placed in a
multi-layer magnetic shield equipped with a system of magnetic
coils within the inner shield that are used to compensate residual
magnetic fields (to a level of $\sim 0.1\ \mu$G), and to apply
well-controlled, arbitrarily-directed fields to the cell
\cite{ourULTRA,ourSens}. The cell is placed between a polarizer
and an analyzer oriented at $\approx 45^{\circ}$ with respect to
each other (a balanced polarimeter). The frequency of the laser
(tuned near the D1-line) is modulated with an amplitude typically
of a few hundred MHz. For the measurements presented here, we have
chosen modulation frequencies up to $\Omega_m = 2 \pi \times 3\
{\rm kHz}$ achieved by modulating a piezo actuator \cite{EN0} in
the diode laser (New Focus, Vortex 6000).
\begin{comment}
We have also carried out measurements where laser frequency was
modulated by changing the diode laser current at frequencies up to
$\sim 100\ {\rm kHz}$. (Modulation at frequencies up to $100\ {\rm
MHz}$ is possible with this laser.) At low modulation frequencies,
results for both type of laser modulation are similar, although
for a quantitative analysis, it is necessary to take into account
non-negligible light amplitude modulation which accompanies
frequency modulation when current modulation is used.
\end{comment}
The amplitudes at the first and second harmonics of $\Omega_m$ of
the difference signal between the two photodiodes at the outputs
of the analyzer are detected with a lock-in amplifier (Stanford
Research Systems, SR810DSP). Upon normalization by the
time-averaged sum of the photodiode outputs, this represents the
signal measured in this experiment (i.e. the effective amplitude
of the first or second harmonic of the time-dependent polarization
rotation angle).

\begin{figure}
\includegraphics[scale=.5]{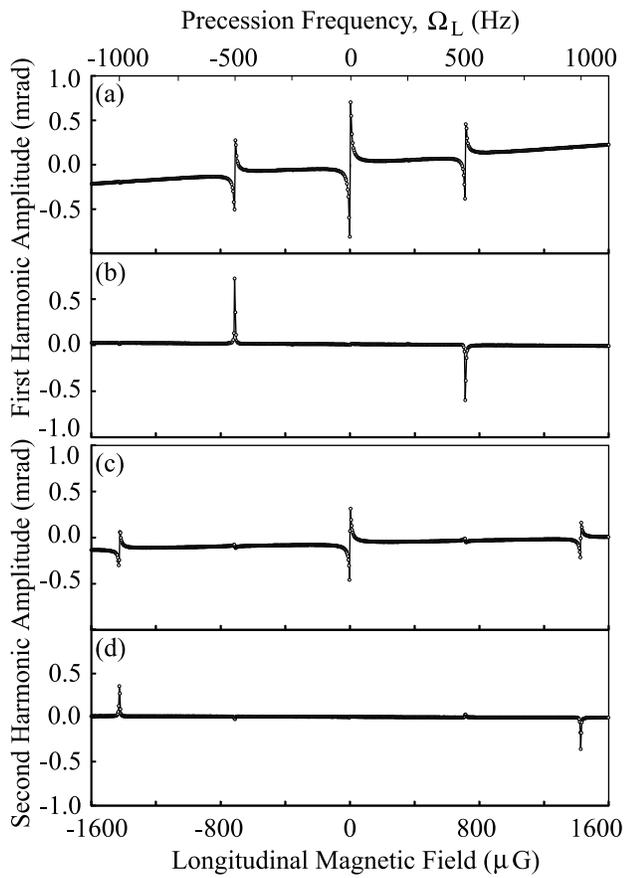}
\caption{Signals detected at the first harmonic (a,b) and second
harmonic (c,d) of $\Omega_m$ as a function of longitudinal
magnetic field. The laser power was $15\ \mu$W, beam diameter
$\sim 2\ $mm, $\Omega_m=2\pi\times1\ $kHz, $\Delta \omega =
2\pi\times220\ $MHz. Traces (a,c) and (b,d) correspond to the
in-phase and the quadrature outputs of the signals from the
lock-in detector, respectively.} \label{FM_SvsB}
\end{figure}

\begin{figure}
\includegraphics[scale=.5]{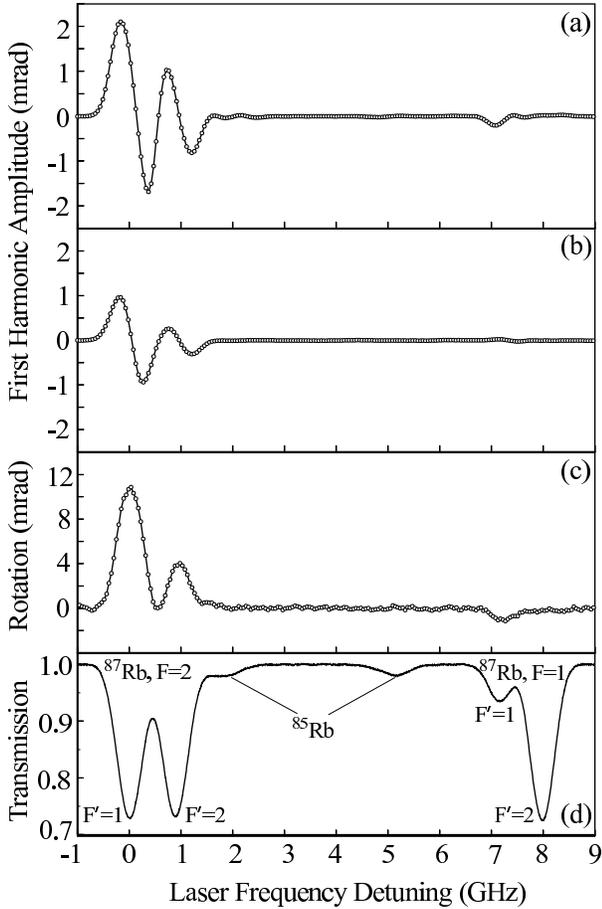}
\caption{Spectral dependences of various signals. (a) First
harmonic amplitude measured at $B=2.2\ \mu$G (the narrow
zero-field resonance), light power $\sim 15\ \mu$W; (b) same at
$B=714.4+2.2\ \mu$G (the $\Omega_m=2 \Omega_L$ resonance), same
light power, $\Omega_m=2\pi\times1\ $kHz, $\Delta \omega =
2\pi\times220\ $MHz; (c) nonlinear Faraday rotation recorded
without frequency modulation (as in \cite{ourULTRA,ourSens}),
$B=2.2\ \mu$G, same light power; (d) transmission through the
vapor cell for low light power ($0.4\ \mu$W). For all traces,
light beam diameter $\sim 2\ $mm.} \label{FM_SvsNu}
\end{figure}

The signals at the first and second harmonics of $\Omega_m$ as a
function of longitudinal (along the light propagation direction)
magnetic field are shown in Fig. \ref{FM_SvsB}. The laser
frequency is modulated at $\Omega_m = 2 \pi \times 1\ \rm{kHz}$
with modulation amplitude $\Delta \omega = 2\pi \times 220\
\rm{MHz}$. For the first and the second harmonic signals, the
laser is tuned to where the respective signals are of maximal
size: the low-frequency slope of the $F=2\rightarrow F'=1$
resonance ($F,F'$ are the total angular momenta of the lower and
upper state) for the first harmonic (Fig. \ref{FM_SvsNu}(a)), and
to the center of the $F=2\rightarrow F'=1$ resonance for the
second harmonic.

For both harmonics, a narrow, dispersively-shaped resonance is
seen in Fig. \ref{FM_SvsB} near $B=0$. Its origin is similar to
that of the ultra-narrow resonances observed in the previous work
\cite{ourULTRA}: the atoms are pumped by linearly-polarized light
into a state aligned \cite{EN1}
\begin{comment}
Alignment is a particular case of atomic polarization
corresponding to an anisotropy characterized by a preferred axis,
rather than by a preferred direction as in the case of
orientation. Alignment is represented by a rank-two (quadrupole)
irreducible tensor component of the density matrix, while
orientation is represented by a rank-one component.
\end{comment}
along the light polarization; optically-pumped atomic alignment
evolves due to the applied magnetic field; the resulting atomic
polarization \cite{EN2}
\begin{comment}
The resulting polarization is alignment rotated due to Larmor
precession at low light powers, and a more complicated atomic
polarization state when optical pumping is saturated
\cite{ourAOC}.
\end{comment}
is ``probed" by the light, leading to a modified light
polarization at the output of the cell. For the zero-field
resonances in Fig. \ref{FM_SvsB}, the modulation frequency is much
faster than the Larmor frequency and the optical pumping rate for
the cell, so that frequency modulation is equivalent to spectral
broadening of the pumping light, which does not significantly
affect the pumping process. On the other hand, modulation is
essential at the probing stage. As the laser frequency is scanned
through the atomic resonance, there is a time-dependent optical
rotation, so the signal contains various harmonics of $\Omega_m$.
The peaks in the magnetic field dependence of the signal occur at
$\Omega_L \equiv g_{F} \mu B = \pm \gamma_{\rm rel}/2$, where
$g_{F}$ is the Land$\acute{\rm e}$ factor ($g_{F}=1/2$ for the
$F=2$ state of $^{87}$Rb), $\mu\approx 2\pi\cdot 1.40 {\rm~MHz/G}$
is the Bohr magneton, and $\gamma_{\rm rel}$ ($\sim 2\pi \cdot
1{\rm~Hz}$ for the data in Fig. \ref{FM_SvsB}) is the relaxation
rate of atomic polarization. The zero-field resonances can be
applied to low-field NMOR-magnetometry \cite{ourSens}, with all of
the advantages of the frequency-modulation technique as in the
case of linear optical rotation \cite{Bar78,Bar88}.

\begin{figure}
\includegraphics[scale=.5]{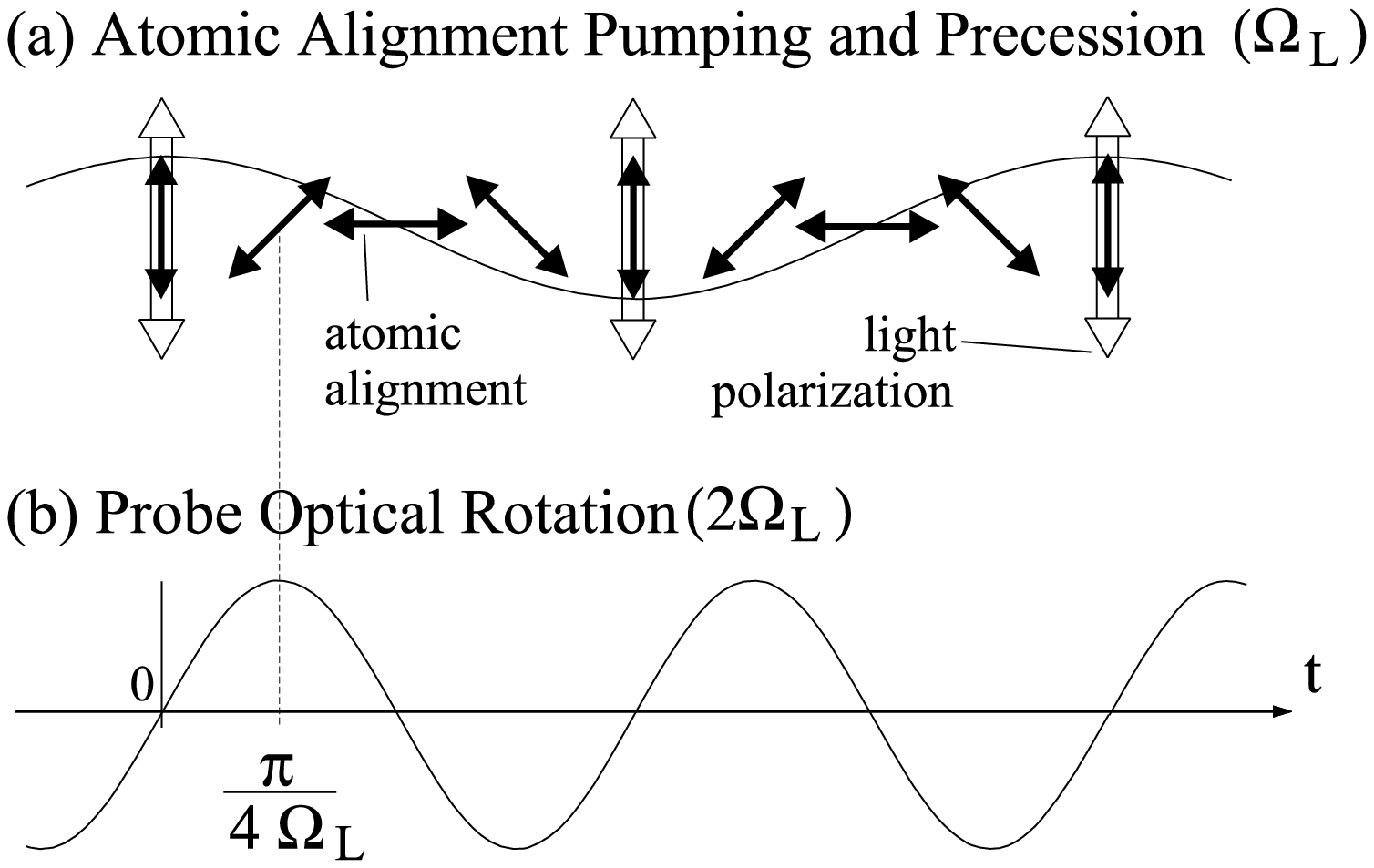}
\caption{Origin of the NMOR resonances at
$\Omega_m=2\Omega_L,\Omega_L$. (a) Atomic alignment precesses with
Larmor frequency (line). Linearly-polarized light periodically
interacts with the atoms. (b) Optical rotation of an unmodulated
probe beam occurs at frequency $2\Omega_L$. } \label{FM_ResExpl}
\end{figure}

Additional resonances are seen at magnetic field magnitudes of
$714.4\ \mu$G ($\Omega_m=2\Omega_L$) and $1428.9\ \mu$G
($\Omega_m=\Omega_L$) for the first and second harmonics,
respectively (Fig. \ref{FM_SvsB}). For these resonances, there are
both dispersively-shaped in-phase signals (Fig.
\ref{FM_SvsB}(a,c)) and $\pi/2$ out of phase (quadrature)
components peaked at the centers of these resonances (Fig.
\ref{FM_SvsB}(b,d)). The origin of the resonances can be
understood as follows (Fig. \ref{FM_ResExpl}). As the laser
frequency is modulated, the optical pumping rate changes depending
on the instantaneous detuning of the laser from the atomic
transition. This occurs with periodicity dictated by the
modulation frequency $\Omega_m$. On the other hand, the state of
atomic polarization also changes periodically due to the presence
of the magnetic field. In the case of alignment transverse to the
magnetic field, this change occurs at frequency $2 \Omega_L$
(alignment precesses at $\Omega_L$, but it returns to the same
state after a half of the corresponding period). If the optical
pumping rate is synchronized with Larmor precession, a resonance
occurs, and the atomic medium is optically pumped into an aligned
state whose axis rotates at $\Omega_L$. The optical properties of
the medium are thus being modulated at $2\Omega_L$, causing
rotation of the light polarization at this frequency. This leads
to dispersively-shaped resonances in optical rotation that are
detected as the signal at the appropriate harmonic of $\Omega_m$.
The quadrature components (Fig. \ref{FM_SvsB}(b,d)) arise because
exactly on resonance, the aligned atoms produce maximum optical
rotation when the alignment axis is at an angle of $\pi/4$ to the
direction of the light polarization. (There is no quadrature
component at zero magnetic field because atomic alignment does not
rotate at all in this case.) Since the widths of the additional
resonances at $\Omega_m=2\Omega_L$ and $\Omega_m=\Omega_L$ are
determined by $\gamma_{\rm rel}$ as for the zero-field resonance,
they are similarly narrow. In addition to these resonances, there
are also smaller resonances in the first- and second-harmonic
signals, generally occurring for commensurate $\Omega_m$ and
$\Omega_L$. There are several mechanisms that lead to these
resonances, which will be described in detail elsewhere.

The overall slope of the curves in Fig. \ref{FM_SvsB} is due to
the so-called transit effect (see e.g. \cite{ourULTRA} and
references therein), for which the peaks occur at $B = \pm
\gamma_{tr}/2 g_{F} \mu \sim 50\ $mG, where $\gamma_{tr}$ is the
rate of atoms' transit through the laser beam.

The spectral dependences of various signals are shown in Fig.
\ref{FM_SvsNu}. Fig. \ref{FM_SvsNu}(a) depicts the spectrum of the
first harmonic signal obtained when the magnetic field was fixed
at $2.2\ \mu$G (corresponding to maximum signal), and the central
frequency of the laser was scanned across the hyperfine structure
of the Rb D1-line (the transmission spectrum is shown for
comparison in Fig. \ref{FM_SvsNu}(d)). Comparing this spectrum
with the NMOR spectrum obtained without frequency modulation (as
in earlier work \cite{ourULTRA,ourSens}) shown in Fig.
\ref{FM_SvsNu}(c), it is seen that the spectral profile of the
signal with frequency modulation is similar to
$d\varphi(\nu)/d\nu$, where $\varphi(\nu)$ is the NMOR angle and
$\nu$ is the laser frequency. Fig. \ref{FM_SvsNu}(b) shows the
spectrum obtained with the magnetic field of $B=(714.4+2.2)~\mu$G,
corresponding to an $\Omega_m=2\Omega_L$ resonance (Fig.
\ref{FM_SvsB}). The spectral dependence of the quadrature
component of the first harmonic for $B=714.4~\mu$G (not shown) is
similar to that of the in-phase component. The signals acquired at
the second harmonic of $\Omega_m$ (not shown) have spectra
resembling $d^2\varphi(\nu)/d\nu^2$.

We have investigated the dependence of the signals on the
modulation amplitude $\Delta \omega$ and on the light power, and
found similar behavior for the zero-field and the
$\Omega_m=2\Omega_L$ resonances. The maximum first harmonic
amplitude is achieved for modulation amplitude $\Delta \omega$
comparable to the Doppler width of the transition $\sim 300\ $MHz.
The signals are proportional to light power at low powers, peak at
around $30\ \mu$W, and decrease with further power increase due to
light broadening. Subsequent work will include systematic
optimization of all parameters for highest magnetometric
sensitivity, as it was done earlier \cite{ourSens} for low-field
NMOR with conventional polarization-modulation polarimetry.
Additional improvement may be expected if the pump and probe laser
beams are separated, and the modulation method is optimized (e.g.,
a combination of frequency and amplitude modulation). With the
present unoptimized parameters and set-up, we estimate that the
shot-noise-limited sensitivity of the device is $\sim 10^{-11}\
G/\sqrt{Hz}$ (comparable or superior, e.g., to the most sensitive
superconducting quantum interference (SQUID) sensors). Also note
that shot-noise-limited operation in the Earth-field range will be
facilitated by the reduced laser noise at high frequencies, and by
the fact that this method uses resonant optical rotation which
allows suppression of the common mode noise of the two
photodetectors and spurious rotations. Due to the absence of
inertia in magnetic spin-precession, a magnetometer based on this
technique (which employs transverse polarization) could, in
principle, also have a high band width ultimately limited by
electronics.

In conclusion, we have demonstrated a novel magnetometric
technique based on nonlinear optical rotation with
frequency-modulated light. This technique is useful for near
zero-field magnetometry and it also opens a possibility for
ultra-sensitive magnetometers with broad dynamic range that
includes the Earth field range, of importance to many
applications, such as geophysics, magnetic prospecting and
navigation. Possible new applications also include nuclear
magnetic resonance spectroscopy and imaging using weakly-polarized
samples that would not require any additional magnet, except the
Earth itself \cite{Tra2001}. Narrow magneto-optical resonances may
also find applications in atomic tests of fundamental symmetries
\cite{ourTCPFP}. For example, it may be advantageous to carry out
a search for a parity- and time-reversal invariance violating
electric dipole moment of Cs in the presence of a Rb
co-magnetometer in a bias magnetic field. Due to the difference
between the Land$\rm{\acute{e}}$ factors for Rb and Cs, this
avoids ``locking" of their polarizations to each other. In
addition, presence of a bias field eliminates the effects of stray
transverse fields.

The authors are grateful to E. B. Alexandrov for stimulating
discussions. This work is supported by the Office of Naval
Research (grant N00014-97-1-0214). D. B. also wishes to
acknowledge support by an NSF CAREER grant.

\bibliography{Magnetometry}

\begin{thebibliography}{12}
\bibitem[{EN1()}]{EN0}
\bibinfo{note}{\ We have also carried out measurements where laser frequency was
modulated by changing the diode laser current at frequencies up to
$\sim 100\ {\rm kHz}$. (Modulation at frequencies up to $100\ {\rm
MHz}$ is possible with this laser.) At low modulation frequencies,
results for both type of laser modulation are similar, although
for a quantitative analysis, it is necessary to take into account
non-negligible light amplitude modulation which accompanies
frequency modulation when current modulation is used.}
\bibitem[{EN1()}]{EN1}
\bibinfo{note}{\ Alignment is a particular case of atomic polarization
  corresponding to an anisotropy characterized by a preferred axis, rather than
  by a preferred direction as in the case of orientation. Alignment is
  represented by a rank-two (quadrupole) irreducible tensor component of the
  density matrix, while orientation is represented by a rank-one component.}
\end{thebibliography}

%--------------------------------------------------------------------
%--------------------------------------------------------------------
%--------------------------------------------------------------------
%--------------------------------------------------------------------
%--------------------------------------------------------------------
%--------------------------------------------------------------------
%--------------------------------------------------------------------
%--------------------------------------------------------------------

\end{document}